\documentclass[preprint,aps,nofootinbib,floatfix]{revtex4-2}
\pdfoutput=1

\usepackage{amsmath}  
\usepackage{amsbsy,amssymb,amsfonts}
\usepackage{graphicx}
\usepackage{color}
\usepackage{epsf}
\usepackage[utf8x]{inputenc}
\usepackage{hyperref}
\usepackage{multirow}
\usepackage{ulem}
\def\bravert{\egroup\,\vrule\,\bgroup}

\newcommand{\beq}{\begin{equation}}
\newcommand{\eeq}{\end{equation}}
\newcommand{\beqa}{\begin{eqnarray}}
\newcommand{\eeqa}{\end{eqnarray}}
\newcommand{\bea}{\begin{array}}
\newcommand{\eea}{\end{array}}

\newcommand{\bef}{\begin{figure}}
\newcommand{\ef}{\end{figure}}
\newcommand{\bc}{\begin{center}}
\newcommand{\ec}{\end{center}}
\newcommand{\bt}{\begin{table}}
\newcommand{\et}{\end{table}}
\newcommand{\btb}{\begin{tabular}}
\newcommand{\etb}{\end{tabular}}

\newcommand{\au}{{\em a.u.}}

\def\rvac{\left| \rule{0.3cm}{.0cm} \right>}

\def\au{{\it a.u.}}
\begin{document}
\normalem
\title {Candidate Molecules for Next-Generation Searches of Hadronic Charge-Parity Violation}

\vspace*{1cm}

\author{Aur\'elien Marc}
\email{marc.aurelien.am@gmail.com}
\affiliation{Laboratoire de Chimie et Physique Quantiques,
             FeRMI, Universit{\'e} Paul Sabatier Toulouse III,
             118 Route de Narbonne,
             F-31062 Toulouse, France}
\author{Micka\"el Hubert}
\email{mickael.hubert@live.fr}
\affiliation{Institut Polytechnique des Sciences Avancées,
             40 boulevard de la Marquette,
	     31000 Toulouse, France }
\author{Timo Fleig}
\email{timo.fleig@irsamc.ups-tlse.fr}
\affiliation{Laboratoire de Chimie et Physique Quantiques,
             FeRMI, Universit{\'e} Paul Sabatier Toulouse III,
             118 Route de Narbonne, 
             F-31062 Toulouse, France }
\vspace*{1cm}
\date{\today}

\begin{abstract}
We systematically study a set of strongly polar heteronuclear diatomic molecules composed of laser-coolable atoms for their
suitability as sensitive probes of new charge-parity violation in the hadron sector of matter. Using relativistic 
general-excitation-rank configuration interaction theory we single out the molecule francium-silver (FrAg) as the
most promising system in this set and calculate its nuclear Schiff-moment interaction constant to 
$W^\mathrm{FrAg}_{SM}(\mathrm{Fr}) = 30168 \pm 2504\au$ for the target nucleus Fr. Our work includes the development
of system-tailored atomic Gaussian basis sets for the target atom in each respective molecule.  

\end{abstract}

\maketitle
\section{Introduction}
\label{SEC:INTRO}
Electric Dipole Moments (EDM) are low-energy probes \cite{https://doi.org/10.48550/arxiv.2203.08103} very effectively used 
in the search for new sources of charge-parity (${\cal{CP}}$) violation beyond those already implemented into the
Standard Model (SM) of elementary particles through the Cabibbo-Kobayashi-Maskawa (CKM) formalism
\cite{Kobayashi,Pontecorvo}. Complex physical systems like atoms and molecules offer distinct advantages in this search
since the new sources of ${\cal{CP}}$ violation can be magnified by many orders of magnitude 
\cite{EDMsNP_PospelovRitz2005,commins_EDM_1999,sandars_PL1965}. The disadvantage of using a complex system lies in the
multitude of possible underlying ${\cal{CP}}$-violating mechanisms creating the EDM at the atomic scale
\cite{Cairncross_Ye_NatPhys2019}, making multiple measurements on different systems necessary in order to disentangle the 
possible sources.

In judiciously chosen atoms and molecules, however, leptonic ${\cal{CP}}$-violation can be strongly suppressed, making 
these systems sensitive to the hadronic and certain semi-hadronic sources only \cite{Barr_eN-EDM_Atoms_1992}. At the
nuclear energy scale one of the leading manifestations of this type of symmetry breaking is the nuclear Schiff moment
\cite{schiff_nucEDM_1963,spevak_auerbach_flambaum1997}.

The nuclear Schiff moment $S$ scales \cite{spevak_auerbach_flambaum1997,flambaum_ginges2002} roughly as
\begin{equation}
	S \propto \frac{\beta_2\beta_3 Z A^{2/3}}{\Delta E_{\pm}}
\end{equation}
where $\beta_2$ and $\beta_3$ are nuclear quadrupole and octupole deformation parameters, respectively,
$Z$ is the proton number, $A$ is the nucleon number and $\Delta E_{\pm}$ is the energy splitting between
opposite-parity doublets of nuclear states. The atomic or molecular interaction $W$ of the nuclear Schiff moment
scales as \cite{flambaum_ginges2002} (for $s_{1/2}-p_{1/2}$ mixing)
\begin{equation}
  W \propto Z^2\, \frac{4\sqrt{1-Z^2\alpha^2}}{\Gamma(2\sqrt{1-Z^2\alpha^2} +1)^2}\,
  \left(\frac{2ZR_N}{a_B}\right)^{2\sqrt{1-Z^2\alpha^2} -2}
  \label{EQ:SCALING}
\end{equation}
where $\alpha$ is the fine-structure constant, $R_N$ is the respective nuclear radius and $a_B$ is the bohr radius.
Highest sensitivity to new ${\cal{CP}}$-violating hadron physics can thus be achieved by using an atomic nucleus with both
large proton number and large octupole deformation. Indeed, regions with nuclear isotopes fulfilling these conditions
have been identified \cite{Budincevic_2015}.

Sensitivity can be further enhanced by several orders of magnitude using a diatomic molecule built from an atom with a 
target nucleus fulfilling the above conditions, another atom with large electron affinity, and assuring that both of
these atoms are laser coolable to very low temperatures \cite{demille_priv}. In Ref. \cite{Fleig_DeMille_2021} a strong case has been
made to use the diatomic molecule radium-silver (RaAg) which is composed of laser-coolable atoms in the search for new
${\cal{CP}}$-violation of leptonic or semi-leptonic origin. It is thus plausible in the search for new hadronic sources 
to use a similar diatomic molecule, but modified such that the leptonic or semi-leptonic sources are suppressed. This
can be achieved when the respective science state of the molecule is predominantly represented by a 
closed-electron-shell configuration. In that case the nucleon-electron scalar-pseudoscalar and electron electric dipole
moment interactions, for instance, will only appear in higher orders of perturbation theory \cite{Fleig_Jung_Xe_2021}.

In this paper we present a systematic study of the Schiff-moment interaction in six diatomic molecules composed of
laser-coolable atoms combined such that they have a polar bond and an electronically closed-shell ground state near
their equilibrium internuclear configuration.
The highly polarizable target atoms are represented by rubidium (Rb), cesium (Cs), and Fr. As partners
with high electron affinity we choose lithium (Li) and Ag. The former atom is mainly included for establishing trends.
For one of these molecules, FrAg, a large set of data relevant to the assembly and trapping in an ultracold
environment has been presented recently \cite{FrAg_Klos_2022}. 
To the best of our knowledge, however, the present are the first calculations of Schiff-moment interactions in Alkali-Alkali
metal molecules.
 
The paper is structured as follows. In the following section \ref{SEC:THEORY} we briefly review the theory of how we
obtain molecular wavefunctions and how the Schiff-moment interaction is calculated using these wavefunctions. Section
\ref{SEC:APPL} comprises the main body where we discuss basis-set optimization for the different molecules and present
results for the molecular Schiff-moment interaction. In the final section \ref{SEC:CONCL} we draw conclusions from our
findings, discuss the expected impact of the results and mention ongoing and future work.

\section{Theory: Molecular Schiff-Moment Interaction Constant}
\label{SEC:THEORY}

The calculation of the Schiff-moment interaction constant in the present molecules follows the implementation
in Ref. \cite{hubert_fleig_2022}.

The molecular ${\cal{P,T}}$-violating energy shift due to a finite nucleus with assumed nuclear Schiff
moment $S_z$ along the molecular axis is written as \cite{Schiff_PRA_2002}
\begin{equation}
 \Delta \varepsilon_{\text{SM}} = -S_z \frac{3}{B}\, \bigg< \sum\limits_{j=1}^n\, {\hat{z}_j}\, \rho({\bf{r}}_j) \bigg>_{\psi}
 \label{EQ:SCHIFF_EXP_MOL}
\end{equation}
where $B = \int\limits_0^{\infty}\, \rho({\bf{r}}) r^4 dr$,
and $\rho({\bf{r}})$ the nuclear charge density
at electron position coordinate ${\bf{r}}$. For a Gaussian nuclear density with exponent $\zeta$ we established that $B = \frac 3{8\pi\zeta}$ 
in Ref. \cite{hubert_fleig_2022}. 
%
%

The molecular wavefunction is obtained from the zeroth-order 
problem
\begin{equation}
 \hat{H} \big| \psi \big>
        = \varepsilon \big| \psi \big>
\end{equation}
with
\begin{eqnarray}
 \nonumber
 \hat{H} &:=& \hat{H}^{\text{Dirac-Coulomb}}  \\
  &=& \sum\limits^n_j\, \left[ c\, \boldsymbol{\alpha}_j \cdot {\bf{p}}_j + \beta_j c^2
-
  \sum\limits^2_K\, \frac{Z_K}{r_{jK}}{1\!\!1}_4 \right]
+ \sum\limits^n_{k>j}\,
\frac{1}{r_{jk}}{1\!\!1}_4 + V_{KL}
 \label{EQ:HAMILTONIAN_MOL}
\end{eqnarray}
for a diatomic molecule with $n$ electrons,
where $K$ runs over nuclei and $V_{KL}$ is the classical electrostatic potential energy for
the two Born-Oppenheimer-fixed nuclei. The CI expansion of the electronic wavefunction reads
\begin{equation}
	\psi\, \widehat{=}\, \left| \Omega \right> = \sum\limits_{I=1}^{{\rm{dim}}{\cal{F}}^t(M,n)}\,
                                       c_{(\Omega),I}\, ({\cal{S}}{\overline{\cal{T}}})_{(\Omega),I} \rvac
        \label{EQ:MOL_WF}
\end{equation}
where $\rvac$ is the true vacuum state,
${\cal{F}}^t(M,n)$ is the symmetry-restricted sector of Fock space ($M_J$ subspace) with $n$ electrons in
$M$ four-spinors,
${\cal{S}} = a^{\dagger}_i a^{\dagger}_j a^{\dagger}_k \ldots$ is a string of spinor creation operators,
${\overline{\cal{T}}} = a^{\dagger}_{\overline l} a^{\dagger}_{\overline m} a^{\dagger}_{\overline n} \ldots$
is a string of creation operators of time-reversal transformed spinors. The determinant expansion coefficients
$c_{(\Omega),I}$ are generally obtained as described in refs. \cite{fleig_gasci,fleig_gasci2}
where $\Omega$ is the total angular momentum projection onto the internuclear axis. 
The Schiff-moment interaction constant for a target
nucleus $A$ of a molecule is then written as
\begin{equation}
        W_{\text{SM}}(A) := \frac{\Delta \varepsilon_{\text{SM}}(A)}{S_z(A)}
                    = -\frac{3}{B}\, \bigg< \sum\limits_{j=1}^n\, {\hat{z}_j}\, \rho_A({\bf{r}}_j) \bigg>_{\psi}.
 \label{EQ:MOLECULAR_SCHIFF_CONSTANT}
\end{equation}
In practical applications $A$ is placed at the origin of the reference frame. The KRCI module of the DIRAC program package 
\cite{DIRAC_JCP} is used for the evaluation of required property integrals \cite{knecht_luciparII,knecht_thesis}.
\section{Schiff-Moment Interactions in Alkali-Lithium and Alkali-Silver Molecules}
\label{SEC:APPL}

\subsection{Technical Details}
\label{Technical Details}

All the numerical calculations presented here were made using a locally modified version of the DIRAC program package \cite{DIRAC_JCP}.
The correlated calculations were carried out through the Configuration Interaction (CI) method with the KRCI module \cite{knecht_luciparII}.
We also use a spin-orbital singles and doubles coupled cluster method with perturbational triples corrections (CCSD(T)) {\it{via}} the
RELCCSD module \cite{visscher_ucc} for the geometry optimization of some of the the studied molecules.

For the target atoms Rb, Cs and Fr we use Dyall's quadruple-zeta (QZ) sets with 
all correlating and dipole-polarizing functions for shells $3d$ through $5s,5p$ added to the primitive set for Rb,
all valence- and available outer-core correlating functions from $n=4$ onward added for Cs, 
and all dipole polarization functions and outer-core correlating functions from $n=5$ onward added for Fr \cite{dyall_s-basis}.
For Cs and Fr we also used the corresponding Dyall's double-zeta (DZ) and triple-zeta (TZ) sets for comparative purposes
\cite{dyall_s-basis}. For Li the EMSL basis sets of cc-pVNZ-DK type with $N \in \{2,\ldots,4\}$ are
employed \cite{EMSL-basis2019}. The Ag atom is described by the same QZ basis set as used in Ref.
\cite{Fleig_DeMille_2021}.

To read this paper, one should know the system of notation we use for the different CI models. The general form here is 
$\mathrm{S}i\_\mathrm{SD}j\_\mathrm{SDT}k\_x\mathrm{a.u.}$ This means that we have $i$ electrons in shells from which 
single excitations are performed, $j$ electrons in accumulated shells with single and double excitations and $k$ electrons 
in accumulated shells with single, double and triple excitations.
$x$ stands for the energy at which we truncate the complementary space.

\subsection{Geometry Optimization}
\label{Geometry Optimization}

In order to have a more accurate interpretation of future experiments, ${\cal{P,T}}$-odd interactions should be evaluated at the 
equilibrium internuclear distance $R_e$ for each of the diatomic molecules. Vibrational fluctuations of these molecular
constants are much smaller than other uncertainties in our calculations.



For the RbLi and FrLi molecules we find the energy-curve minimum by fitting a polynomial to respectively $9$ (RbLi) and $7$ (FrLi) calculated 
CCSD(T) data points in a range from $5.8$ \au\ to $6.7$ \au\ (RbLi) and $6.5$ \au\ to $7.1$ \au\ (FrLi), respectively.
The results obtained are shown in table (\ref{Tbl:Re-6-Molecules}), along with literature values for the other four molecules. 
For RbLi our result agrees with literature results 
\cite{allouche_rbli,igel_mann_alkali,urban_sadlej_alkali} to within 1.5\%. For FrLi a very recent Fock-space CC calculation 
\cite{Lamberti_LiFr_2023} using relativistic pseudopotentials deviates from our result by less than $1$\%.

\begin{table}[h] 
	\centering
	\caption{$R_e$ from CCSD(T) calculation; $^*$value taken from RCCSD(T) calculations in Ref. \cite{Smialkowski_Tomza_2021}; $^{\dagger}$ Dirac-Coulomb CCSD(T) with 22 active electrons \cite{Soerensen_spinfree}}
	\label{Tbl:Re-6-Molecules}
	\begin{tabular}{l|c|c|c|c|c|c}
		Molecule   & RbLi & CsLi & FrLi & RbAg & CsAg & FrAg\\ \hline
		$R_e$ [a.u.] &  6.527 & 6.927$^{\dagger}$ & 6.878 & 5.845$^*$ & 6.112$^*$ & 6.190$^*$ 
	\end{tabular}
\end{table}

\subsection{Basis-Set Optimization}
\label{Basis-Set Optimization}

In this section we will demonstrate the necessity to do a basis-set optimization and discuss how it is carried out.

\subsubsection{Deficiency of standard basis sets}
\label{Deficiency of standard basis sets}

\begin{table} 
	\centering
	\caption{$W_{SM}$ with double-, triple- and quadruple-Zeta basis set at the DCHF level for RbLi, CsLi and FrAg}
	\label{Table:CsLi/FrAg_standard_sets}
	\begin{tabular}{l|cc|cc|cc}
		& 
		\multicolumn{2}{c|}{RbLi ($R_e = 6.527 \au$)} & 
		\multicolumn{2}{c|}{CsLi ($R_e = 6.927 \au$)} & 
		\multicolumn{2}{c}{FrAg ($R_e = 6.190 \au$)} \\ \hline
		Basis & 
		\multicolumn{1}{c|}{ $\varepsilon_{\text{DCHF}}$ [a.u.]} & $W_{\text{SM}}$ [a.u.] & 
		\multicolumn{1}{c|}{ $\varepsilon_{\text{DCHF}}$ [a.u.]} & $W_{\text{SM}}$ [a.u.] & 
		\multicolumn{1}{c|}{ $\varepsilon_{\text{DCHF}}$ [a.u.]} & $W_{\text{SM}}$ [a.u.]   
		\\ \hline
		cvDZ &
		\multicolumn{1}{c|}{ -2987.2228737} & -1829.3 & 
		\multicolumn{1}{c|}{ -7794.1925854} & -10110.1 & 
		\multicolumn{1}{c|}{-29622.7980959} &  5946      
		\\ 
		cvTZ & 
		\multicolumn{1}{c|}{ -2987.2366325} & -2144.5 & 
		\multicolumn{1}{c|}{ -7794.2033064} & -2848.8 & 
		\multicolumn{1}{c|}{-29622.8345496} &  28173    
		\\
		cvQZ & 
		\multicolumn{1}{c|}{ -2987.2370600} & -1150.7 & 
		\multicolumn{1}{c|}{ -7794.2038442} &  2098.1 & 
		\multicolumn{1}{c|}{-29622.8362766} &  29451    
		\\
	\end{tabular}
\end{table}

The results in Table \ref{Table:CsLi/FrAg_standard_sets} demonstrate that small atomic basis sets (cvDZ) yield spurious results for $W_{SM}$, confirming earlier findings \cite{hubert_fleig_2022}. Even though the cvQZ basis produce a result physically acceptable for CsLi and FrAg molecules, it is not converged yet. Furthermore, the cvQZ basis 
produces a physically unacceptable result for the RbLi molecule. 
Therefore, even a cvQZ basis can fail to describe the physics of the Schiff-moment interaction in a qualitatively correct manner. We thus optimized target-atom basis sets in the molecular framework for each of the considered molecules.


\subsubsection{Basis set optimization itself}
\label{Basis set optimization itself}

We follow the procedure proposed in reference \cite{hubert_fleig_2022}. This densification procedure allows us to generate a basis 
set that resembles quintuple-zeta (5Z) quality in the absence of a fully-optimized 5Z basis set for the target atom. However, we 
have to determine individually when to truncate the densification procedure. This is done at the DCHF level considering both 
convergence of $W_{SM}$ as well as stability of total energy.

\begin{table} 
	\centering
	\caption{1-densified-cvQZ basis-set determination : DCHF for $^1\Sigma_0$ for 3 Alkali-Li molecules}
	\label{Table:Alkali-Li_densification}
	\begin{tabular}{l|cc|cc|cc}
		& 
		\multicolumn{2}{c|}{RbLi ($R_e =  6.527 \au$)} & 
		\multicolumn{2}{c|}{CsLi ($R_e = 6.927 \au$)} & 
		\multicolumn{2}{c}{FrLi ($R_e = 6.878 \au$)} \\ \hline
		Basis & 
		\multicolumn{1}{c|}{ $\varepsilon_{\text{DCHF}}$ [a.u.]} & $W_{\text{SM}}$ [a.u.] & 
		\multicolumn{1}{c|}{ $\varepsilon_{\text{DCHF}}$ [a.u.]} & $W_{\text{SM}}$ [a.u.] & 
		\multicolumn{1}{c|}{ $\varepsilon_{\text{DCHF}}$ [a.u.]} & $W_{\text{SM}}$ [a.u.]   
		\\ \hline
		cvQZ &
		\multicolumn{1}{c|}{-2987.2370600} & -1150.7 & 
		\multicolumn{1}{c|}{-7794.2038442 } & 2098 & 
		\multicolumn{1}{c|}{-24315.6247270} & 24004 
		\\ 
		d-cvQZ+0sp & 
		\multicolumn{1}{c|}{-2987.2370672} & -1183.4 & 
		\multicolumn{1}{c|}{-7794.2038120} & 1761 & 
		\multicolumn{1}{c|}{-24315.6219173} & 242340 
		\\
		d-cvQZ+1sp & 
		\multicolumn{1}{c|}{-2987.2370734} & -414.6 & 
		\multicolumn{1}{c|}{-7794.2038394} & 2887 & 
		\multicolumn{1}{c|}{-24315.6237748} & 25434   
		\\
		d-cvQZ+2sp & 
		\multicolumn{1}{c|}{-2987.2370747} & 389.9 & 
		\multicolumn{1}{c|}{-7794.2038393} & 2885 & 
		\multicolumn{1}{c|}{-24315.6237716} & 25393   
		\\
		d-cvQZ+3sp & 
		\multicolumn{1}{c|}{-2987.2370749} & 808.1 & 
		\multicolumn{1}{c|}{-7794.2038393} & 2891 & 
		\multicolumn{1}{c|}{-24315.6237716} & 25328   
		\\
		d-cvQZ+4sp & 
		\multicolumn{1}{c|}{-2987.2370752} & 844.5 & 
		\multicolumn{1}{c|}{-7794.2038393} & 2883 & 
		\multicolumn{1}{c|}{-24315.6237714} & 25233   
		\\
		d-cvQZ+5sp & 
		\multicolumn{1}{c|}{-2987.2370751} & 842.0 & 
		\multicolumn{1}{c|}{-7794.2038393} & 2894 & 
		\multicolumn{1}{c|}{-24315.6237716} & 25341   
		\\
		d-cvQZ+6sp &
		\multicolumn{1}{c|}{-2987.2370750} & 842.8 & 
		\multicolumn{1}{c|}{-7794.2038393} & 2884 & 
		\multicolumn{1}{c|}{} &    
		\\
		d-cvQZ+7sp &
		\multicolumn{1}{c|}{-2987.2370750} & 842.3 & 
		\multicolumn{1}{c|}{-7794.2038393} & 2887 & 
		\multicolumn{1}{c|}{} &    
		\\
		d-cvQZ+8sp &
		\multicolumn{1}{c|}{-2987.2370750} & 843.0 & 
		\multicolumn{1}{c|}{-7794.2038393} & 2888 & 
		\multicolumn{1}{c|}{} &    
		\\
	\end{tabular}
\end{table}

For the RbLi and RbAg molecules (Tables \ref{Table:Alkali-Li_densification} and \ref{Table:Alkali-Ag_densification}), we can see a 4\% change when going from 3sp to 4sp and, respectively, 0.3\% and 0.1\% when going from 4sp to 5sp. Meanwhile the energy is converged at $10^{-6}\au$ Thus we chose a d-cvQZ+4sp augmented Dyall QZ basis that we denote cvQZ+ for the following when concerning those molecules.

For the CsLi and CsAg molecules (Tables \ref{Table:Alkali-Li_densification} and \ref{Table:Alkali-Ag_densification}), $W_{SM}$ changes when going from 1sp to 2sp level by less than 0.1\% while the energy is converged at $10^{-6}\au$ In the following, cvQZ+ will be a short version of d-cvQZ+1sp when referring to those molecules.

For the FrLi and FrAg molecules (Tables \ref{Table:Alkali-Li_densification} and \ref{Table:Alkali-Ag_densification}) we observe one order of magnitude difference for $W_{SM}$ between 0sp and other densification levels because the basis is not complete enough to describe the physics of the interaction. However the $W_{SM}$ change when going from 1sp to 2sp is, respectively, about 0.2\% and 0.7\%. In addition, according to the convergence at $10^{-5}\au$ of the energy, the d-cvQZ+1sp basis is an optimal 1-densified basis and the one we will use and denote as cvQZ+ for those molecules. 

\begin{table} 
	\centering
	\caption{1-densified-cvQZ basis-set determination : DCHF for $^1\Sigma_0$ for 3 Alkali-Ag molecules}
	\label{Table:Alkali-Ag_densification}
	\begin{tabular}{l|cc|cc|cc}
		& 
		\multicolumn{2}{c|}{RbAg ($R_e = 5.845 \au$)} & 
		\multicolumn{2}{c|}{CsAg ($R_e = 6.112 \au$)} & 
		\multicolumn{2}{c}{FrAg ($R_e = 6.190 \au$)} 
		\\ \hline
		Basis & 
		\multicolumn{1}{c|}{ $\varepsilon_{\text{DCHF}}$ [a.u.]} & $W_{\text{SM}}$ [a.u.] & 
		\multicolumn{1}{c|}{ $\varepsilon_{\text{DCHF}}$ [a.u.]} & $W_{\text{SM}}$ [a.u.] & 
		\multicolumn{1}{c|}{ $\varepsilon_{\text{DCHF}}$ [a.u.]} & $W_{\text{SM}}$ [a.u.]   
		\\ \hline
		cvQZ &
		\multicolumn{1}{c|}{-8294.4487693} & -1445.2 & 
		\multicolumn{1}{c|}{-13101.4159681} & 2594.2 & 
		\multicolumn{1}{c|}{-29622.8363749} & 29475    
		\\ 
		d-cvQZ+0sp & 
		\multicolumn{1}{c|}{-8294.4487772} & -1313.8 & 
		\multicolumn{1}{c|}{-13101.4159361} & 2742.2 & 
		\multicolumn{1}{c|}{-29622.8335672} & 193210   
		\\
		d-cvQZ+1sp & 
		\multicolumn{1}{c|}{-8294.4487834} & -493.3  & 
		\multicolumn{1}{c|}{-13101.4159635} & 3593.8 & 
		\multicolumn{1}{c|}{-29622.8354237} & 31349    
		\\
		d-cvQZ+2sp & 
		\multicolumn{1}{c|}{-8294.4487846} & 486.5   & 
		\multicolumn{1}{c|}{-13101.4159634} & 3589.2 & 
		\multicolumn{1}{c|}{-29622.8354202} & 31143    
		\\
		d-cvQZ+3sp & 
		\multicolumn{1}{c|}{-8294.4487848} & 1015.6  & 
		\multicolumn{1}{c|}{-13101.4159635} & 3580.5 & 
		\multicolumn{1}{c|}{-29622.8354203} & 31085    
		\\
		d-cvQZ+4sp & 
		\multicolumn{1}{c|}{-8294.4487850} & 1058.7  & 
		\multicolumn{1}{c|}{-13101.4159634} & 3576.2 & 
		\multicolumn{1}{c|}{-29622.8354201} & 31039    
		\\
		d-cvQZ+5sp & 
		\multicolumn{1}{c|}{-8294.4487849} & 1057.2  & 
		\multicolumn{1}{c|}{-13101.4159634} & 3585.1 & 
		\multicolumn{1}{c|}{-29622.8354203} & 31094    
		\\
		d-cvQZ+6sp &
		\multicolumn{1}{c|}{-8294.4487849} & 1059.2 & 
		\multicolumn{1}{c|}{} &  & 
		\multicolumn{1}{c|}{} &    
		\\
		d-cvQZ+7sp &
		\multicolumn{1}{c|}{-8294.4487849} & 1057.5 & 
		\multicolumn{1}{c|}{} &  & 
		\multicolumn{1}{c|}{} &    
		\\
		d-cvQZ+8sp &
		\multicolumn{1}{c|}{-8294.4487849} & 1059.0 & 
		\multicolumn{1}{c|}{} &  & 
		\multicolumn{1}{c|}{} &    
		\\
	\end{tabular}
\end{table}

\subsection{Correlated Calculations}
\label{Correlated Calculations}

Now that we have all the optimized basis sets for our six molecules, we present and discuss results including inter-electron correlation
effects obtained using the CI method with various models.

\subsubsection{Alkali Atoms bound to Li}
\label{Alkali Atoms bound to Li}

\begin{table} 
	\centering
	\caption{ RbLi, $^1\Sigma_0$, $R_e =$ 6.527 a.u. 
	}
	\label{Table : RbLi W_SM}
	\begin{tabular}{l|r}
		Basis/cutoff                &  $W_{\text{SM}}$ [a.u.] \\ \hline
		cvQZ+/DCHF                  &  $844.5$      \\
		cvQZ+/SD2\_9.9au            &  $819.1$      \\
		cvQZ+/SD10\_9.9au           &  $823.8$      \\
		cvQZ+/SD22\_9.9au           &  $814.0$      \\
	\end{tabular}
\end{table}

\begin{table} 
	\centering
	\caption[]{ CsLi, $^1\Sigma_0$, $R_e =$ 6.927 a.u. 
	}
	\label{Table : CsLi W_SM}
	\begin{tabular}{l|r}
		Basis/cutoff                &  $W_{\text{SM}}$ [a.u.] \\ \hline
		cvQZ+/DCHF                  &  $ 2886.9$      \\
		cvQZ+/SD10\_10au            &  $ 2795.1$      \\
		cvQZ+/SD8\_SDT10\_10au      &  $ 2844.2$      \\
		cvQZ+/SD22\_10au            &  $ 2813.4$      \\
	\end{tabular}
\end{table}

\begin{table} 
	\centering
	\caption[]{ FrLi, $^1\Sigma_0$, $R_e =$ 6.878 a.u. 
	}
	\label{Table : FrLi W_SM}
	\begin{tabular}{l|r}
		Basis/cutoff                &  $W_{\text{SM}}$ [a.u.] \\ \hline
		cvQZ+/DCHF                  &  $25434$      \\
		cvQZ+/SD2\_2au              &  $24288$      \\
		cvQZ+/SD2\_5au              &  $24282$      \\
		cvQZ+/SD2\_10au             &  $24287$      \\
		cvQZ+/SD10\_10au            &  $24245$      \\
		cvQZ+/S10\_SD22\_10au       &  $24289$      \\
		cvQZ+/SD22\_10au            &  $24414$      \\
	\end{tabular}
\end{table}

In tables \ref{Table : RbLi W_SM},\ref{Table : CsLi W_SM} and \ref{Table : FrLi W_SM} are shown the results from our calculations on the different Alkali-Li diatomic molecules. These include various CI models and show the corresponding value for $W_{SM}$. 

First of all, general trends for electron correlation effects established in Ref. \cite{hubert_fleig_2022} are also observable in the
present systems: Including excitations out of shells that directly contribute to $s-p$ mixing diminishes the interaction constant, and this
comprises the principal correlation effect in all studied molecules.
Furthermore, from these tables we can directly see that the heavier the target atom, the higher the molecular Schiff-moment interaction constant.
Indeed, with equivalent models for the 3 molecules we obtained $W_{SM}^{RbLi}(\mathrm{Rb, SD22\_9.9a.u.}) = 814 \au$, $W_{SM}^{CsLi}(\mathrm{Cs, SD22\_10a.u.}) = 2813 \au$ and $W_{SM}^{FrLi}(\mathrm{Fr, SD22\_10a.u.}) = 24414 \au$ The models are equivalent because they correlate the valence shell and the $(n-1)s$, $(n-1)p$ and $(n-2)d$ target atom shells and the $2s$ Li shell. 
There is roughly a factor of 3.5 difference between RbLi and CsLi and a factor of 30 difference between RbLi and FrLi. Such a trend was expected 
and confirms the scaling of the Schiff-moment interaction given in Eq. \ref{EQ:SCALING}.

Francium is thus a very good choice of target atom for a Schiff-moment sensitive molecule since the resulting molecular constant is 8.3 times 
larger than with cesium.

\subsubsection{Alkali Atoms bound to Ag}
\label{Alkali Atoms bound to Ag}

\begin{table} 
	\centering
	\caption[]{RbAg, $^1\Sigma_0$, $R_e =$ 5.845 a.u. 
	}
	\label{Table : RbAg W_SM}
	\begin{tabular}{l|r}
		Basis/cutoff               &  $W_{\text{SM}}$ [a.u.] \\ \hline
		cvQZ+/DCHF                 &  $ 1059.1$      \\
		cvQZ+/SD2\_2au             &  $ 1036.8$      \\
		cvQZ+/SD2\_6.8au           &  $ 1036.7$      \\
		cvQZ+/SD2\_11au            &  $ 1036.8$      \\
		cvQZ+/S10\_SD12\_11au      &  $ 1015.0$      \\
		cvQZ+/SD12\_11au           &  $ 1041.9$      \\
	\end{tabular}
\end{table}

\begin{table} 
	\centering
	\caption[]{CsAg $^1\Sigma_0$, $R_e =$ 6.112 a.u. 
	}
	\label{Table : CsAg W_SM}
	\begin{tabular}{l|r}
		Basis/cutoff              &  $W_{\text{SM}}$ [a.u.] \\ \hline
		cvQZ+/DCHF                &  $3593.4$      \\
		cvQZ+/SD2\_2au            &  $3503.8$      \\
		cvQZ+/SD2\_11au           &  $3502.3$      \\
		cvQZ+/S10\_SD12\_2au      &  $3446.6$      \\
		cvQZ+/S10\_SD12\_5.5au    &  $3423.5$      \\
		cvQZ+/S10\_SD12\_11au     &  $3420.9$      \\
		cvQZ+/SD12\_11au          &  $3529.6$      \\
	\end{tabular}
\end{table}

\begin{table} 
	\centering
	\caption[]{FrAg, $^1\Sigma_0$, $R_e =$ 6.190 a.u. 
	}
	\label{Table : FrAg W_SM}
	\begin{tabular}{l|c|r}
		Basis/cutoff            &  $\varepsilon_{\text{CI}}$ [a.u.] &$W_{\text{SM}}$ [a.u.] \\ 
		\hline
		cvQZ+/DCHF	         & -29622.8354238 &  $31350$      \\ 
		cvQZ+/SD2\_2au       & -29622.8604657 &  $30359$      \\ 
		cvQZ+/SD2\_3au       & -29622.8605116 &  $30349$      \\ 
		cvQZ+/SD2\_5au       & -29622.8605445 &  $30355$      \\ 
		cvQZ+/SD2\_8au       & -29622.8605500 &  $30360$      \\ 
		\hline
		cvQZ+/SD10\_8au      & -29623.0196812 &  $29980$      \\ 
		cvQZ+/SDT10\_8au     & -29623.0260848 &  $29909$      \\ 
		\hline
		cvQZ+/SD12\_8au      & -29623.1920759 &  $30711$      \\ 
		\hline
		cvQZ+/SD20\_8au      & -29623.3371101 &  $30127$      \\ 
		\hline
		cvQZ+/SD36\_5au      & -29623.7102434 &  $30333$      \\ 
		cvQZ+/SD36\_8au      & -29623.8379481 &  $30239$      \\ 
	\end{tabular}
\end{table}

In Tables \ref{Table : RbAg W_SM}, \ref{Table : CsAg W_SM} and \ref{Table : FrAg W_SM}  are shown the results from our calculations on the different Alkali-Ag diatomic molecules. These include various CI models and their corresponding values for $W_{SM}$.

First of all we can compare the Alkali-Ag diatomic molecule with the corresponding Alkali-Li diatomic molecule. 
At the valence-shell level, the $W_{SM}^{Alkali-Ag}(\mathrm{Alkali})$ value is roughly 1.25 times larger than the corresponding $W_{SM}^{Alkali-Li}(\mathrm{Alkali})$ one. This can largely be explained by the difference in electron affinity (EA) between Ag and Li \cite{EA_Li,Bilodeau_Scheer_Haugen_1998}.
Indeed, since the EA of the silver atom is twice as big as the lithium EA, 
the partial negative charge forming on the polarizing atom partner is significantly greater when Ag is used instead of Li.
This axial distortion of the electron cloud is accompanied by mixing of predominantly $s$ and $p$ spinors and consequently leads to differences in the Schiff moment interaction.

So far we have shown that the francium atom is a good target atom and that the silver atom is a good perturber atom. 
Thus we mainly focus on the FrAg molecule in the following analysis and interpretation. 

In addition to Schiff-moment interaction constants we display total CI energies in Table \ref{Table : FrAg W_SM} to show the 
magnitude of correlation energies in the various shells and to verify that our calculations are in accord with the variation 
theorem of quantum mechanics. Our results for $W_{SM}^{FrAg}(\mathrm{Fr})$ show the expected pattern for electron 
correlation effects. Replacements out of the outermost
shells that contribute directly to $s-p$ mixing reduce the Schiff-moment interaction (CI models SD2 and SD10). This principal effect
amounts to about $-4.5$\%. In line with this interpretation, 
double excitations from the $4d$ Ag shell (model SD12) lead to a increase by about $1$\% relative to the model SD2.
If correlations between the $4d$ Ag and the $6s6p$ Fr electrons are taken into account (model SD20) this increase is roughly halved.


Adding the $5d$ Fr and $4p$ Ag shells to all of the above comprises the model SD36 which increases $W_{SM}^{FrAg}(\mathrm{Fr})$ by about
0.4\% relative to the model SD20 and gives the value of 30239 a.u. at an $8\au$ cutoff in the complementary space. 
In view of these results the following core shells are expected to give only minor corrections due to large energy
denominators. In addition to this, compensations between effects that increase and those that quench $W_{SM}^{FrAg}(\mathrm{Fr})$
will further diminish the additional corrections not taken into account in our explicit models. 

Concerning the truncation of virtual spinors with the model SD36 we observe a difference between cutoff at 5 a.u. and
cutoff at 8 a.u. of only 0.3\% in $W_{SM}^{FrAg}(\mathrm{Fr})$. 
Since the effect due to this change in cutoff is of the same magnitude as the change between the two most encompassing correlation models,
SD20 and SD36, we consider it unnecessary to include virtual spinors of higher energy in the wavefunction expansion.

As we have demonstrated at the CISD level around $80$\% of total correlation effects are contributed by the valence and $6s6p$ Fr shells.
Thus, it is of interest to look at a higher-level model to the correlations arising from these electrons. 
The model SDT10 adds the full set of triple excitations to the model SD10 and increases the drop in $W_{SM}^{FrAg}(\mathrm{Fr})$ from
the SD10 model by about $7$\%, indicating that even higher excitations should be rather unimportant.

In order to obtain a final value for $W_{SM}^{FrAg}(\mathrm{Fr})$ we use the model SD36 as a basis to which we add a triples correction:
$$W_{SM}^{FrAg}(\mathrm{Fr},\mathrm{SD}36\_8\mathrm{a.u.}) - W_{SM}^{FrAg}(\mathrm{Fr},\mathrm{SD}10\_8\mathrm{a.u.}) + W_{SM}^{FrAg}(\mathrm{Fr},\mathrm{SDT}10\_8\mathrm{a.u.}) = 30168 \au$$

To this final value we attribute an uncertainty of 8.3\% coming from the different physical approximations and models used. 6.4 parts in
these 8.3\% are attributed to basis-set incompleteness, 0.9 parts to the correlation models used (cutoff, number of electrons correlated and 
excitation rank) and $1$ part is attributed to the physical approximation in the Hamiltonian.
Thus, the final value we obtain including its uncertainty is $W^\mathrm{FrAg}_{SM}(\mathrm{Fr}) = 30168 \pm 2504\au$, assuming that
further uncertainty associated with the operator \cite{Schiff_PRA_2002} describing the interaction of the Schiff moment with the electron 
shells is negligible.

\section{Conclusions and Outlook}
\label{SEC:CONCL}

We establish in this work the FrAg diatomic molecule as an excellent probe in the search for new sources of hadronic ${\cal{CP}}$
violation. The use of a Ag atom as polarizing partner increases the molecular Schiff-moment interaction by nearly $25$\% as 
compared to the more standard \cite{salomon_licool_2013} and also significantly electron-affine Li atom as polarizing partner in 
the FrLi molecule.

FrAg consists of laser-coolable atoms that form a strongly polar molecular bond and exhibit a molecular Schiff-moment 
interaction that is only roughly $25$\% weaker than same interaction in the thallium monofluoride (TlF) molecule \cite{CENTREX_2021},
the molecule currently in use in a leading hadron-sector ${\cal{CP}}$-violation search.
According to our DCHF calculations of the respective ground states of TlF and FrAg the partial negative charge on F in TlF is 
significantly greater than on Ag in FrAg. Alongside this greater polarization the amount of $s-p$ mixing in TlF is also 
greater than in FrAg in the outermost valence spinors, largely explaining the difference in Schiff-moment interaction constant.
Nevertheless, using laser-coolable atoms \cite{Fleig_DeMille_2021} instead of a molecular beam and a target nucleus with strong
octupole deformation should greatly outweigh this rather modest disadvantage.

In ongoing work we are exploring systematically the effect of replacing the Gaussian nuclear density used for Schiff-moment
interactions thus far by a more accurate Fermi distribution.
Moreover, we are studying an important semi-hadronic EDM source, the nucleon-electron tensor-pseudotensor (Ne-TPT) 
interaction in scientifically relevant molecules using the atomic basis sets optimized in the present work.


\section{Acknowledgments}
\label{SEC:ACK}
We thank David DeMille (Chicago) for helpful comments.
\bibliographystyle{unsrt}
\newcommand{\Aa}[0]{Aa}

\clearpage
\end{document}